\documentclass[fleqn,usenatbib]{mnras}

\usepackage{newtxtext,newtxmath}

\usepackage[T1]{fontenc}

\DeclareRobustCommand{\VAN}[3]{#2}
\let\VANthebibliography\thebibliography
\def\thebibliography{\DeclareRobustCommand{\VAN}[3]{##3}\VANthebibliography}

\usepackage{graphicx}	% Including figure files
\usepackage{amsmath}	% Advanced maths commands
\usepackage{CJK}

\title[Chaotic Type I Migration]{Chaotic Type I Migration in Turbulent Discs}

\author[Wu, Chen \& Lin., 2023]{
Yinhao Wu (吴寅昊),$^{1}$\thanks{\href{mailto:email@domain}{yw505@leicester.ac.uk} (YW), \href{mailto:email@domain}{yc9993@princeton.edu} (YC)}
Yi-Xian Chen (陈逸贤),$^{2}$\footnote[1]~
Douglas N. C. Lin (林潮)$^{3,4}$
\\
$^{1}$School of Physics and Astronomy, University of Leicester, Leicester LE1 7RH, UK\\
$^{2}$Department of Astrophysical Sciences, Princeton University, USA\\
$^{3}$Department of Astronomy \& Astrophysics, University of California, Santa Cruz, CA 95064, USA\\
$^{4}$Institute for Advanced Studies, Tsinghua University, Beĳing 100084, China
}

\date{Accepted XXX. Received YYY; in original form ZZZ}

\pubyear{2023}

\begin{document}
\begin{CJK*}{UTF8}{gbsn} 
\label{firstpage}
\pagerange{\pageref{firstpage}--\pageref{lastpage}}
\maketitle

\begin{abstract}
By performing global hydrodynamical simulations of accretion discs with driven turbulence models, 
we demonstrate that elevated levels of turbulence induce highly stochastic migration torques on low-mass companions embedded in these discs. 
This scenario applies to planets migrating within gravito-turbulent regions of protoplanetary discs as well as stars and black holes embedded in the outskirts of active galactic nuclei (AGN) accretion discs. 
When the turbulence level is low, linear Lindblad torques persists in the background of stochastic forces and its accumulative effect can still dominate over relatively long timescales. However, 
in the presence of very stronger turbulence, classical flow patterns around the companion embedded in the disc are disrupted, 
leading to significant deviations from the expectations of classical Type I migration theory over arbitrarily long timescales. Our findings suggest that the stochastic nature of turbulent migration can prevent low-mass companions from monotonically settling into universal migration traps within the traditional laminar disc framework, 
thus reducing the frequency of three-body interactions and hierarchical mergers compared to previously expected. 
We propose a scaling for the transition mass ratio from classical to chaotic migration $q\propto \alpha_R$, where $\alpha_R$ is the Reynolds viscosity stress parameter, which can be further tested and refined by conducting extensive simulations over the relevant parameter space.
%The chaotic migration of companions in accretion discs has been a focal point of theoretical research, offering vital insights into disc evolution and planetary formation. This complex phenomenon has wide-ranging implications, including the predictability of migration traps and subsequent celestial interactions. In this study, we explore the implications of turbulent forces on the migration behavior of planets and sub-stellar black holes (sBHs) in protoplanetary discs (PPDs) and active galactic nuclei (AGN) discs, respectively. 
%Using hydrodynamical simulations with driven turbulence models, we demonstrate that high levels of disc turbulence lead to highly stochastic driving forces on migrating bodies. For a low-mass companion, low turbulence shows that classical Lindblad and corotation torques dominate over time, whereas stronger turbulence disrupts traditional flow patterns, deviating from classical Type I torque expectations. Our findings suggest that turbulent migration can prevent objects like sBHs from settling into migration traps, thus reducing the likelihood of frequent three-body interactions and hierarchical mergers.
\end{abstract}


%%%%%%%%%%%%%%%%%%%%%%%%%%%%%%%%%%%%%%%%%%%%%%%%%%

%%%%%%%%%%%%%%%%% BODY OF PAPER %%%%%%%%%%%%%%%%%%

% ======================================= %
\section{Introduction}\label{sec:intro}
% ======================================= %

Planetary migration 
occurs within the early stages of planet formation, 
when planets tidally interact with their protoplanetary discs (PPDs). 
Type I migration primarily affects 
low-mass planets\citep{Goldreich1980, Ward1997}, typically those up to a few times the mass 
of Earth, when the companion lacks the mass required to 
create a gap in the disc density profile \citep{LinPapaloizou1986}. 
It plays an important role in the formation of super Earths
and cores of proto-gas-giant planets \citep{ida2008, benz2014, liu2015}.
Early analytic analysis\citep{Ward1997} and numerical simulations \citep{Paardekooper2010}
of this process primarily focus on the determination of tidal torque between the
disc and planets with fixed orbits.  The pace and direction of this migration 
are determined by the sum of the differential (usually negative) Lindblad and 
unsaturated (often positive) corotation torque.  
The details of torque imbalance is set by the disc structure 
\citep{Paardekooper2010,Kley2012}, particularly the temperature and surface density 
gradients. Type I migration usually leads to inward motion and occurs 
much faster than both the growth of giant planet cores and the evaporation 
of the disc \citep{Ward1997,Tanaka2002}. However, depending on the specific
heating mechanisms and accretion-disc structures, the total torque can exhibit 
local positivity in certain regions of the PPD. This in turn can cause 
stalling of Type I migration among a range of planetary mass \citep{Kretke2012}, 
forming effective migration traps at the boundary of outward migration regions\citep{liu2015}.
The potentially rapid migration speed warrants self consistent simulations
in which the embedded planet's orbit evolve together with the disc structure.
Moreover, recent studies \citep{Wafflard-Fernandez_2020,
Wu_2023} show that observational signatures generated by migrating planets 
may differ from those produced by stationary planets. 

Stellar-mass Black Holes (sBHs) also migrate and evolve in active galactic nuclei (AGN) accretion discs around Supermassive Black Holes (SMBHs). The hierarchical mergers of sBHs could possibly contribute to LIGO-Virgo gravitational wave (GW) events \citep{McKernan2014,McKernan2018}. 
Early population studies of these sBHs and their GW features invoke classical Type I migration formulae to account for their orbital decay due to tidal interaction with the disc, 
driving them to converge radially towards certain migration traps existent in the AGN disc structure \citep{Bellovary2016} and undergo frequent mergers and interactions \citep{Tagawa2020}.

Turbulence ubiquitously exist in both kinds of discs, albeit at different levels. 
In mid-planes of PPDs, Magneto-rotational Instability (MRI) turbulence level is proven to be low,
with an amplitude $\alpha \lesssim 10^{-3}$ in the $\alpha$ presciption \citep{shakura1973}
in numerical simulations \citep{2013BaiStone}, due to low ionization level and non-ideal MHD effects. 
Such upper limits are also confirmed by observations of molecular line broadening \citep{Flaherty2017,Flaherty2020,Giovanni-2023}. 
This level of turbulence have been shown to drive stochastic Type I migration of low-mass planets in short-term MHD simulations \citep{Nelson2005}, 
but it has been pointed out by \citet{BaruteauLin2010} that such levels of turbulence is not able to disrupt the classical planetary wake structure, 
and the residue differential Lindblad torque can still dominate on the long term to drive inward migration. 
This justifies the use of an effective viscosity $\alpha$ in laminar disc simulations
to mimick the effect of turbulent accretion in planet-disc interaction and conclude Type I differential Lindblad torque dependencies, 
as in the widely applied \citet{Paardekooper2010} formalism. 

However, planetesimal or planet core formation might occur very early during PPDs' evolution \citep{Xu2022,Xu2023,Yamato2023,Han2023}.  During their infancy, PPDs are massive and dense.  Marginal gravitation instability
leads to amplification of local disturbance and gravito-turbulence induced effective viscosity corresponds to an average 
effective Reynolds $\langle \alpha_R \rangle$
as large as 0.05 \citep{Gammie2001, Johnson2003} or 0.1\citep{lin1987, deng2020}.
Outskirts of AGN discs are also subject to marginal gravitational instability \citep{paczynski1978},
gravito-turbulence \citep{lin1988}, and intense star formation \citep{Goodman2003,Thompson2005,Jiang2011,Chen2023}. 
With a swarm of embedded massive stars 
co-evolving inside the disc \citep{Cantiello2021,alidib2023, Huang2023}, they generate density waves which 
interfere and dissipate, producing an effective turbulent viscosity whose magnitude
positively correlates with the mass and number density of embedded objects \citep{Goodman2001}. If strong turbulence is able to significantly alter the flow pattern around the companion,
applying linear migration torques in population synthesis of these objects or adding simple stochastic torques on top of a linear component, e.g. in the prediction sBH evolution in AGN discs and GW statistics from their mergers \citep{Secunda2019,Tagawa2020}, may no longer be accurate.

In this letter, we revisit the chaotic migration problem as posed in \citet{BaruteauLin2010}, but focus on what occurs in the highly turbulent regime applicable to the fore-mentioned contexts. 
We emphasise that for $\langle\alpha_R\rangle \gtrsim 0.1$, migration of low-mass companions can indeed become chaotic over long timescales as classical circum-companion flow pattern becomes disrupted. The Letter is organised as follows: in \S \ref{sec:setup}, we introduce our numerical setup for turbulent companion-disc interaction; We will present our results in \S \ref{sec:results} and give a brief conclusion in \S \ref{sec: conclusion}.

% ======================================= %
\section{Numerical Setup}\label{sec:setup}
% ======================================= %

To explore the effect of strong turbulence on Type I migration, 
we apply a modified version of the hydrodynamic grid code FARGO \citep{Masset2000} 
with a phenomenological turbulence prescription that follows \citet{Laughlin2004} and \citet{BaruteauLin2010}. 

The initial and boundary conditions are similar to that adopted by \citet{BaruteauLin2010}. 
The axisymmetric isothermal disc has aspect ratio $h = h_0(r/r_0)^{1/2}$, with $h_0 = 0.03$. 
The initial surface density is $\Sigma = \Sigma_0 (r/r_0)^{-1/2}$, with $\Sigma_0$ a normalization constant irrelevant to the dynamics. 
The embedded companion with a mass of $M_{\rm p}$, or a mass ratio of $q = M_{\rm p}/M_*$ compared to its host mass $M_*$, is initially at the orbital radius of $r_0=1$, with other code unit being $G = M_* = 1$, the orbital frequency $\Omega_0 = \Omega(r_0) = 1$. 
Self-gravity is neglected because turbulence is prescribed and not generated, 
and the instantaneous migration torque that the disc exerts on the planet's orbital angular momentum is normalized in units of $\Sigma_0 r_0^4\Omega_0^{2}$. 
The planet's orbit self-consistently evolves under the influence of the torque, but since the extent of migration is relatively small within a few thousand orbits, the magnitude of torque isn't affected even if we normalize with $\Sigma r^4\Omega^{2}$ at real-time locations. 

The simulation domain covers $(0.4-1.8)r_0$ in the radial direction, and $0-2\pi$ in the azimuthal direction. 
We resolve the disc by $N_r = 512$ radial zones with logarithmic spacing, and $N_\phi = 1536$ azimuthal zones. wave-killing zones are adopted close to the boundaries to minimize unphysical wave reflections \citep{deValBorro2006}. 
To prevent gas velocity from diverging infinitely close to the companion, the planet gravity is softened by a smoothing (Plummer) length of $\epsilon = 0.6h_0 r_0$.

To mimic turbulence, a fluctuating potential $\Phi_{\rm turb} \propto \gamma$ is applied to the disc, 
corresponding to the superposition of 50 wave-like modes \citep{Laughlin2004} such that

\begin{equation}
    \Phi_{\text {turb }}(r, \phi, t)=\gamma r^{2} \Omega^{2} \sum_{k=1}^{50} \Lambda_{k}(m_k, r, \varphi, t)
\end{equation}
where $\gamma$ is the dimensionless characteristic amplitude of turbulence. 
Each stochastic factor for the $k$-th mode, $\Lambda_{k}$, expressed as 

\begin{equation}
    \Lambda_{k} = \xi_k e^{-(r-r_k)^2/\sigma_k^2} \cos(m_k\phi -\phi_k-\Omega_k \tilde{t}_k) \sin (\pi \tilde{t}_k/\Delta t_k),
\end{equation}
is associated with a randomly drawn wavenumber $m_k$ from a logarithmically uniform distribution ranging from $m=1$ to the maximum wave number $m_{\rm max}$. The initial radial $r_k$ and azimuthal $\phi_k$ location of the mode are drawn from
uniform distribution. The modes’ radial extent is $\sigma_k = \pi r_k /4m$.
Modes start at time $t_{0,k}$, and their lifetime is $\Delta t_k = 0.2\pi r_k / m c_s$,
with $c_s$ being the local sound speed. $
\Omega_k$ is the Keplerian frequency at $r_k$, $\tilde{t}_k = t - t_{0,k}$ ,and $\xi_k$
is a dimensionless constant drawn from a Gaussian distribution of unit width. 
As discussed in \citet{BaruteauLin2010}, the choice of parameters for this turbulence driver emulates the power spectrum of a typical Kolmogorov cascade, following the $m^{-5/3}$ scaling law up to $m_{\rm max}$. 
An effective Reynold stress parameter $\langle \alpha_R \rangle$ around the companion location can be calibrated from the velocity fluctuations to be  approximately 
\begin{equation}
    \langle \alpha_R \rangle \simeq 35 (\gamma/h_0)^2,
    \end{equation}
i.e. proportional to the square of the turbulence amplitude (see \citet{BaruteauLin2010} for details).

%The total viscosity controlling vortensity diffusion is $\langle \alpha \rangle\approx 4 \langle \alpha_R \rangle$ 

% ======================================= %
\section{Results}\label{sec:results}
% ======================================= %

% FFFFFFFFFFFFFFFFFFFFFF %
\begin{figure}
\centering
\includegraphics[width=0.99\hsize]{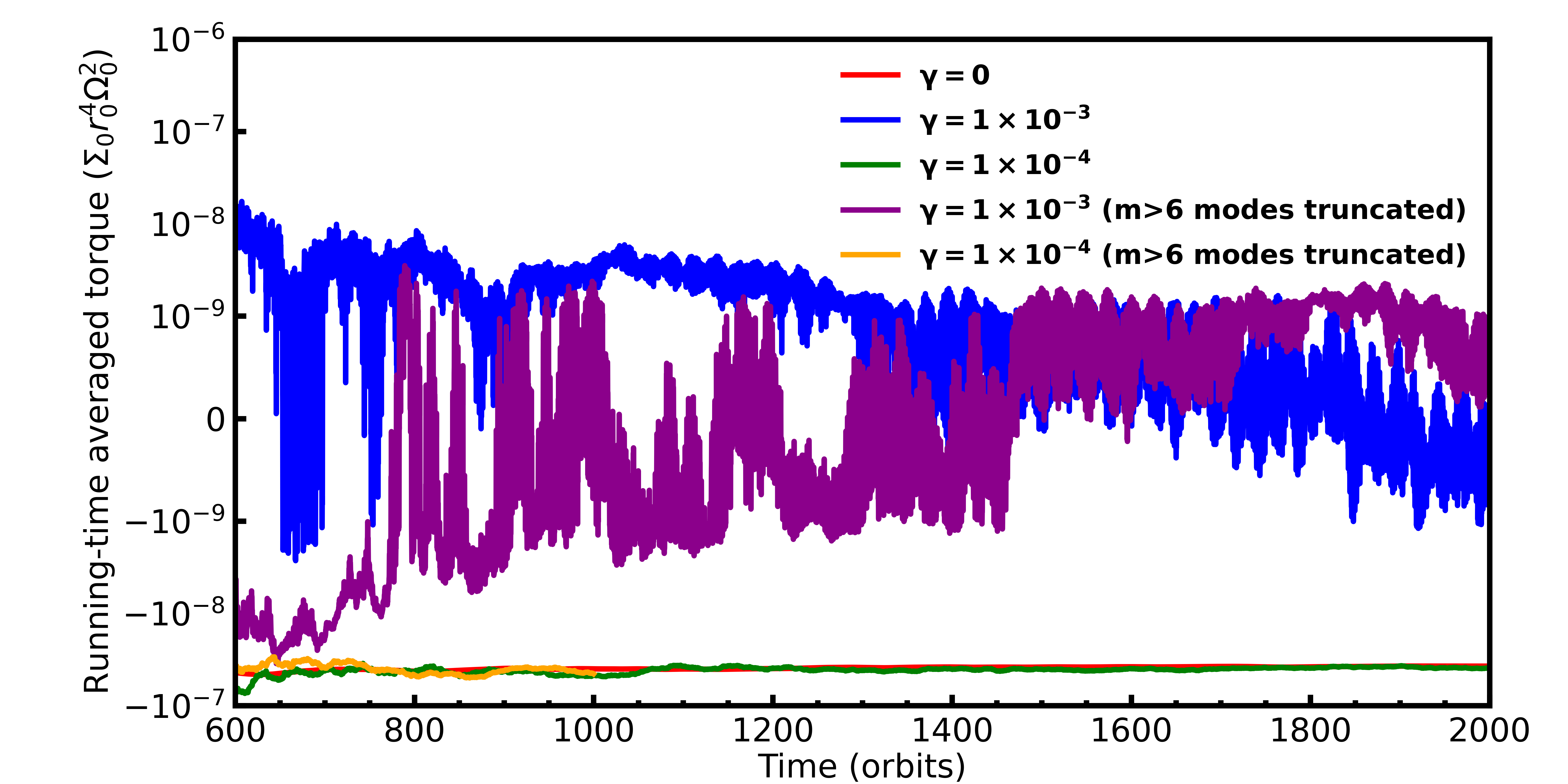}
\caption{Time evolution of the running-time averaged specific torque (normalized by $\Sigma_0 r_0^4 
\Omega_0^2$) for a fixed-orbit companion with a fiducial mass ratio of $q = 5\times10^{-6}$ ($\sim 1.6 M_{\oplus}$) compared to a solar mass host starting from 500 orbits. Different colors represented either runs with different turbulence amplitude $\gamma $, or with different selection of mode azimuthal wavenumbers. Discarding small scale turbulence (truncating $m>6$ modes) does not introduce significant difference in our results. }\label{fig:rtatorque}
\end{figure}
% FFFFFFFFFFFFFFFFFFFFFF %

\subsection{Dependence on Turbulence Strength}

In Figure \ref{fig:rtatorque} we plot the running-time average of migration torques, measured in units of $\Sigma_0 r_0^4 \Omega_0^2$, for a fiducial set of parameters with companion mass ratio $q = 5\times 10^{-6}$ and $h_0 = 0.03$. Different colors represent either runs with different turbulence amplitude $\gamma$ or with different selection of mode azimuthal wavenumbers.

The classical formula \citep{Tanaka2002,Paardekooper2010}
for type I migration torque scaling is
\begin{equation}
    \Gamma = - C_{\rm L} (q/h_0)^2 \Sigma_0 r_0^4 \Omega_0^2.
    \label{eqn:classic}
\end{equation}
Calibrating with the steady-state $\gamma = 0$ result, we obtain $C_{\rm L} \approx 1.8$
(Figure \ref{fig:rtatorque} red line). This quantity is dominated by the Lindblad torque because in 
the inviscid ($\gamma=0$) disc, corotation torque saturates \citep{Goldreich1980}. Although $\gamma = 10^{-4}$
brings in extra turbulence and effective viscosity, the running average of the torque nevertheless converges to the value
corresponding to $C_{\rm L} \approx 1.8$ after a few hundred orbits. By default we include all $m$ up to grid scale, 
but we tested that truncating higher order modes makes very little difference (Figure \ref{fig:rtatorque}, green and yellow lines), as 
the large-scale (low-$m$) modes are most influential to migration.

\begin{figure}
\centering
\includegraphics[width=0.99\hsize]{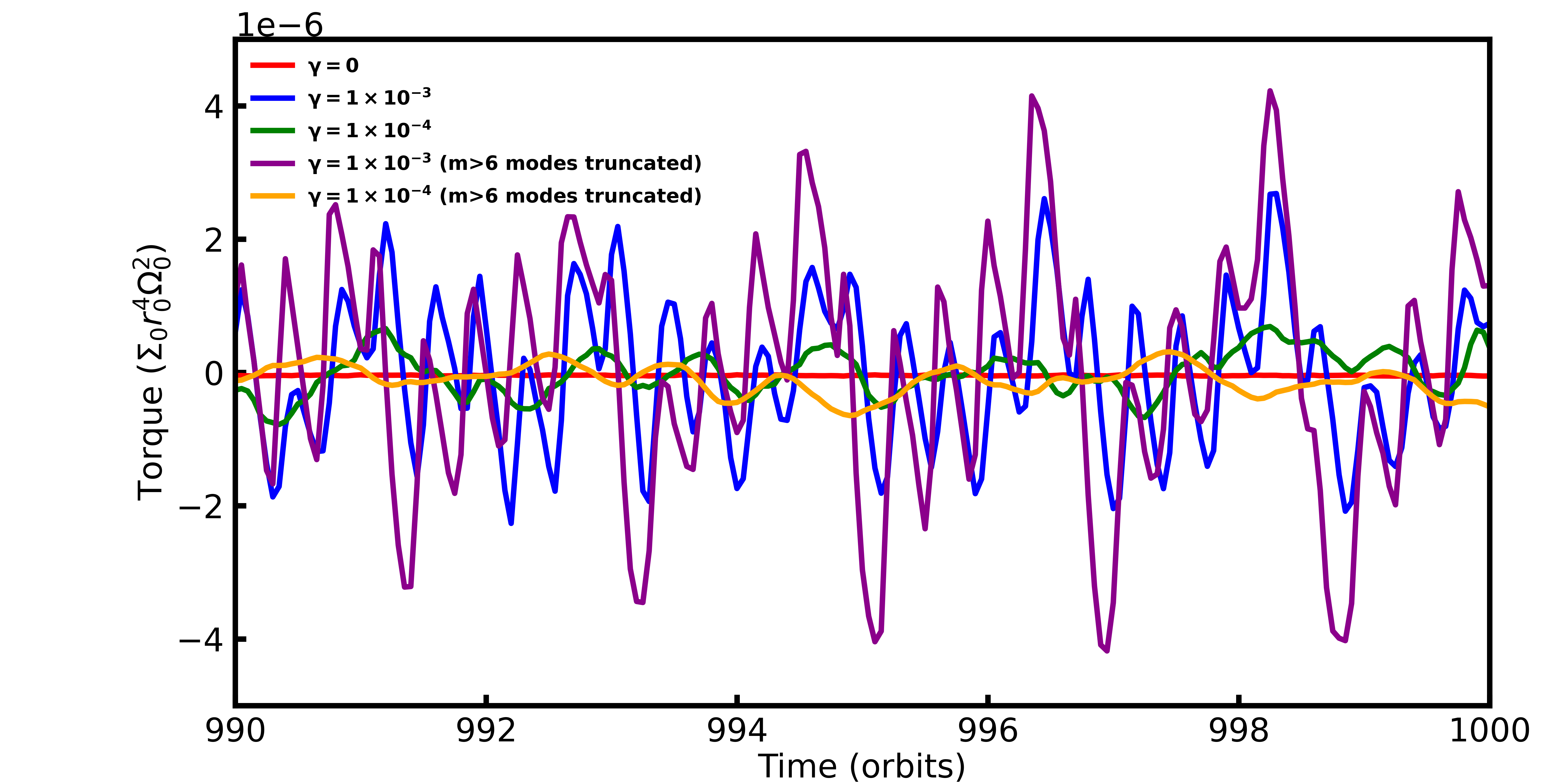}
\includegraphics[width=0.99\hsize]{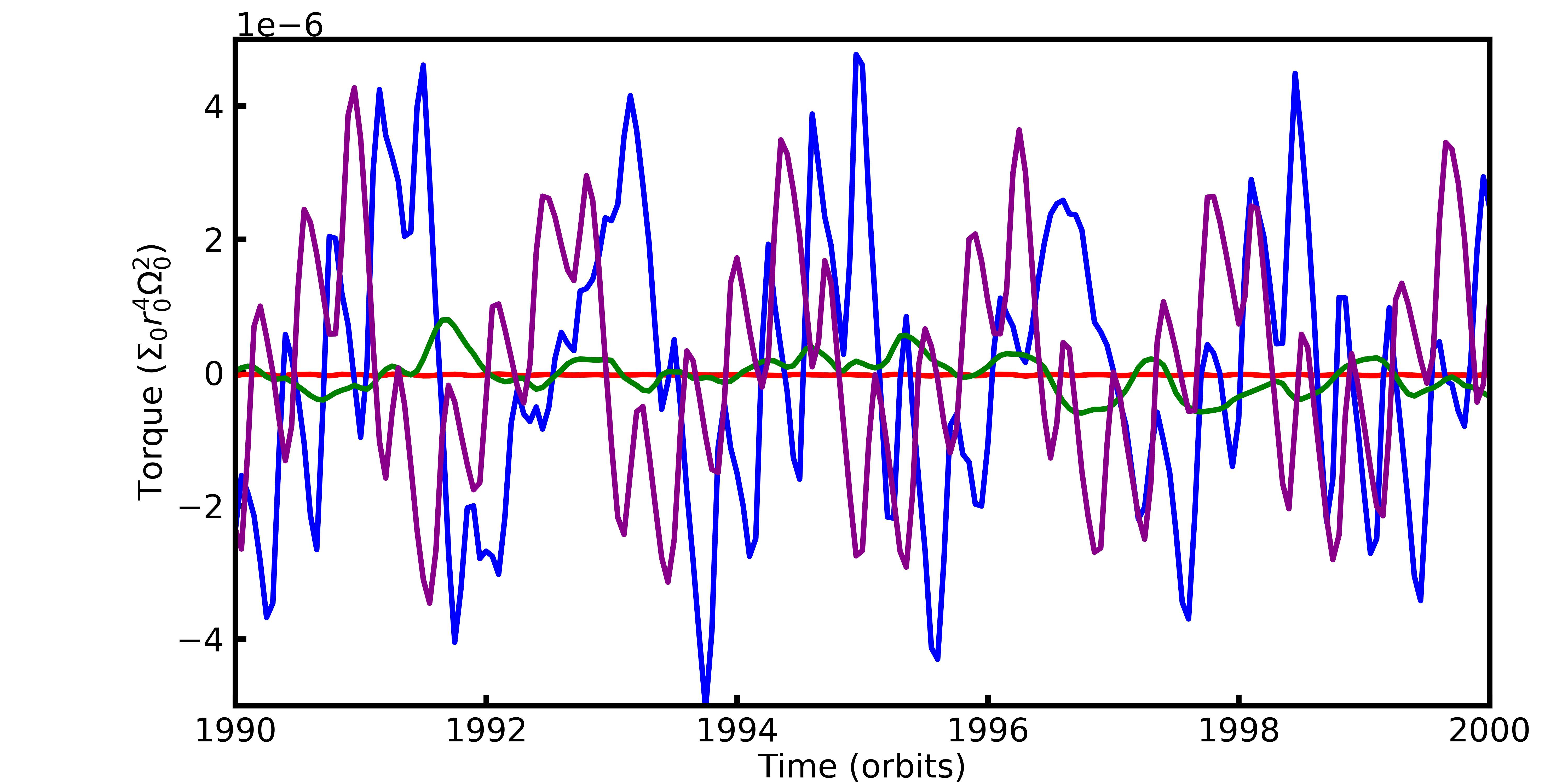}
\caption{The total migration torques in units of $\Sigma_0 r_0^4 \Omega_0^2$ at around 1000 orbits (upper panel) and 2000 orbits (lower panel) respectively for the fiducial case. 20 data points are collected per orbit.}\label{fig:torque}
\end{figure}

Such behaviour is expected from the analysis of \citet{BaruteauLin2010}. 
When turbulence level is moderate, structure of density waves and flow patterns of classical planet-disc interaction can still manifest after averaging the density distribution, therefore the migration torque can be seen as the superposition of a random fluctuating component with magnitude of 

\begin{equation}
\langle  \Gamma \rangle= {1 \over T} \left[ \left( \int_T \Gamma^2 dt \right)-\left( \int_T \Gamma dt \right)^2 \right]^{{1/2}}
= \langle C_{\rm turb} \rangle (q/h_0)^2 \Sigma_0 r_0^4 \Omega_0^2 
\end{equation}
and a continuous component close to the classical type-I torque Eqn \ref{eqn:classic}.  
In a quasi-steady state, the fluctuating amplitude is typically much (by nearly 2 order of magnitude) 
larger than the continuous component as plotted in 
Figure \ref{fig:torque}, while its accumulative effect will decay with time.

\begin{figure}
\centering
\includegraphics[width=0.99\hsize]{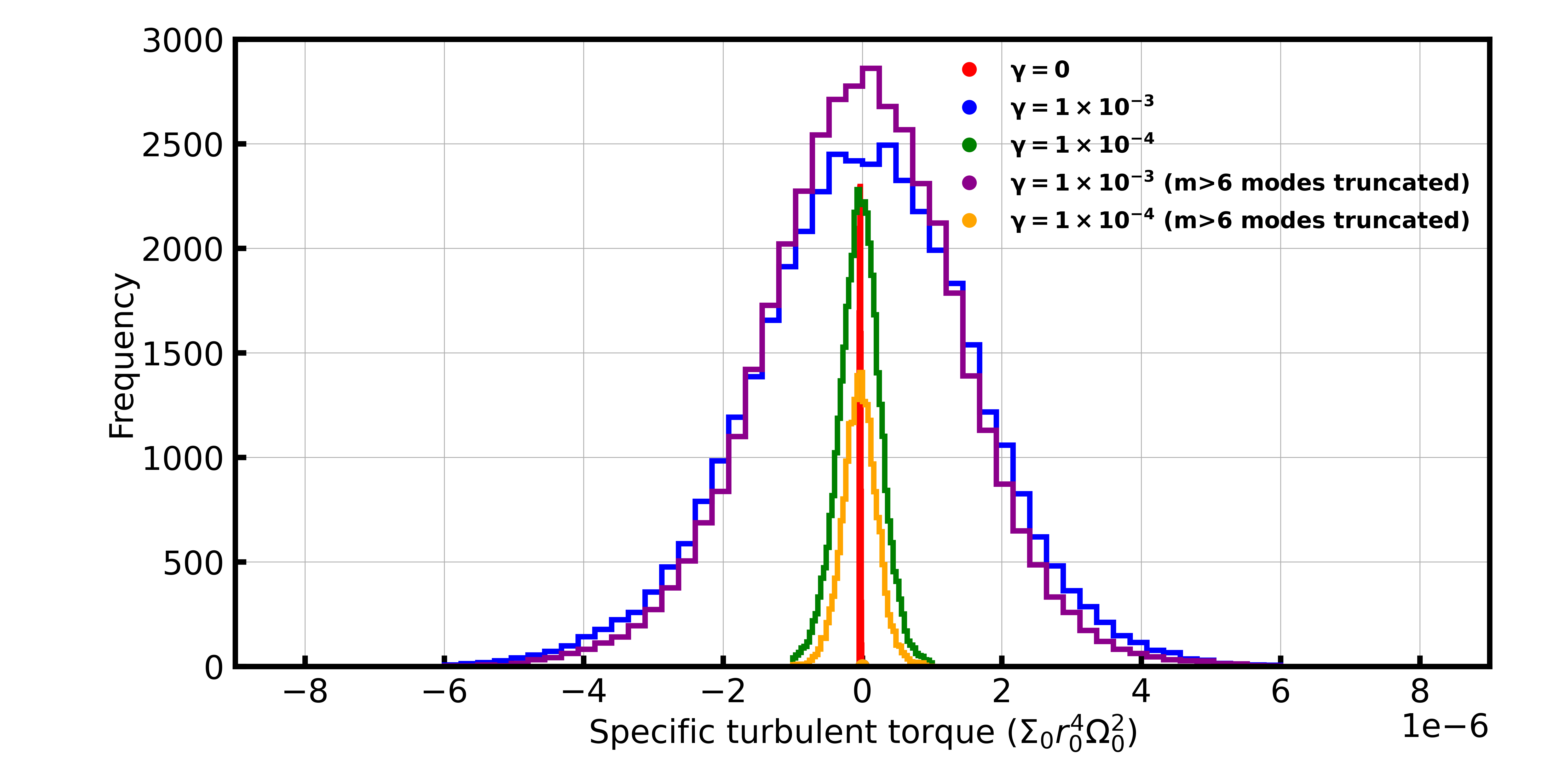}
\caption{Distribution of the specific turbulent torque obtained from the  $q=5\times10^{-6}$ results. The total number of data samples 
is 30000 (20 data points per orbit from 500 to 2000 orbits). We take 50 bins between $[-6\times10^{-6},6\times10^{-6}]$ for the $\gamma = 10^{-3}$ cases, between $[-1\times10^{-6},1\times10^{-6}]$ for the $\gamma = 10^{-4}$ cases, between $[-5\times10^{-8},5\times10^{-8}]$ for the $\gamma = 0$ case to compare the dispersion of the torques. 
}\label{fig:fre}
\end{figure}

With the auto-correlation/eddy-turnover timescale of the turbulent component being 
$\sim$ one planetary orbit, the time for the accumulative effect of the linear torque 
to be significant is $\tau_{\rm conv} \sim (\langle C_{\rm turb}  \rangle/C_{\rm L})^2$ orbits.
From the measured torque data, 
we estimate the standard deviation of Gaussian-distributed torques to be $\langle C_{\rm turb} \rangle = 8$ 
and $63$ for $\gamma=10^{-4}$ and $10^{-3}$ respectively, consistent with the histogram of total torques (Figure \ref{fig:fre}) that show a 
roughly Gaussian dispersion dominated by the fluctuating component with typical dispersion being $\sim \langle C_{\rm turb} \rangle (q/h_0)^2$. 
In our numerical simulations,
the running average torque for $\gamma=10^{-4}$ cases quickly converge with the $\gamma = 0$ scenario within $\ll 100$ orbits (Figure \ref{fig:rtatorque}), 
corresponding to 
the expectation of $\tau_{\rm conv} \sim 20$ orbits.
While for the $\gamma=10^{-3}$ cases, 
the average torque quickly decays to values two orders of magnitude below the linear expectation, instead of showing any trend of settling towards the linear prediction. 

\begin{figure}
\centering
\includegraphics[width=0.99\hsize]{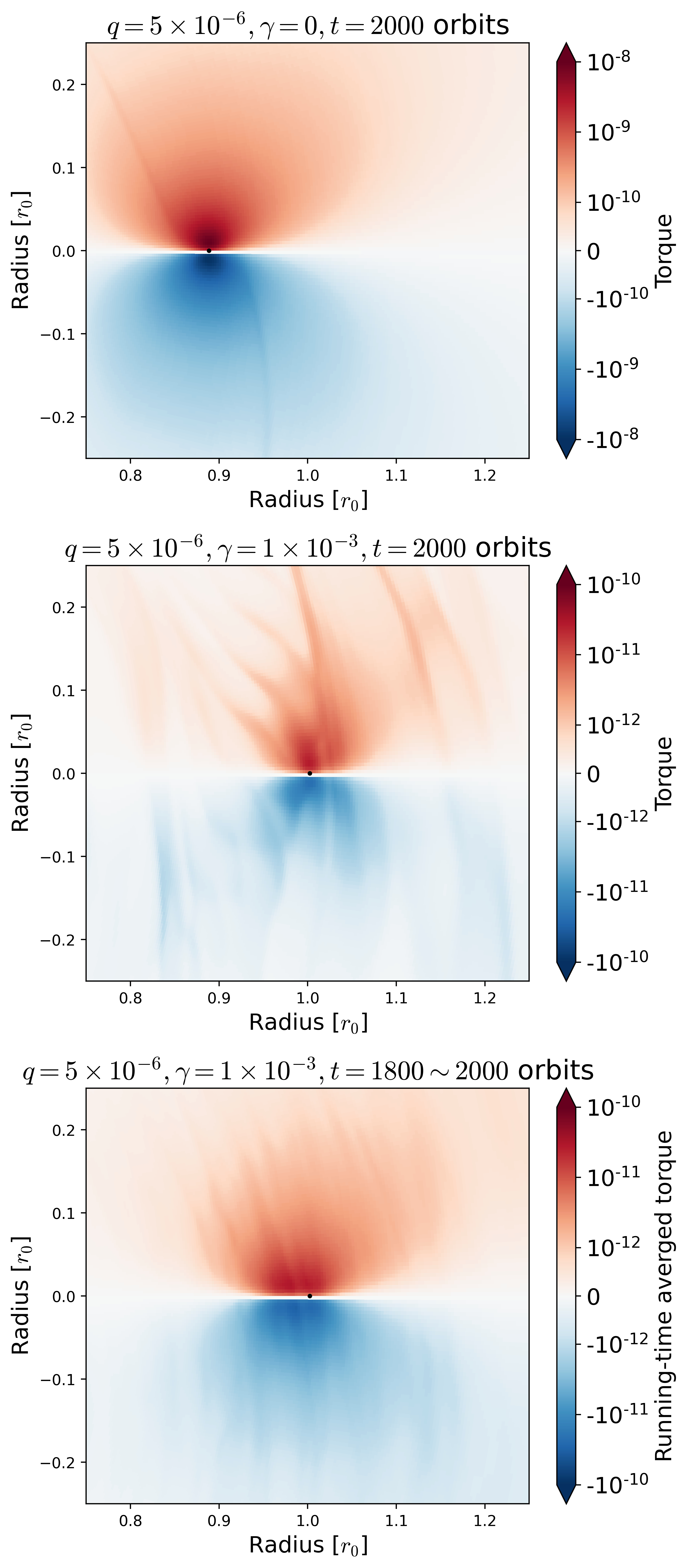}
\caption{Top: the snapshot of migration torque at $t = 2000$ orbits for $q = 5\times10^{-6}$ with $\gamma = 0$. Middle: the snapshot of migration torque at $t = 2000$ orbits for $q = 5\times10^{-6}$ with $\gamma = 1\times10^{-3}$. Bottom: the snapshot of the running-time averaged specific torque for $q = 5\times10^{-6}$ with $\gamma = 1\times10^{-3}$, calculated between $t = [1800,2000]$ orbits. In each panel, the black dot represents the position of the embedded companion. {The quantities are plotted with a Cartesian frame and zoomed in towards the companion region.}}\label{fig:2d}
\end{figure}

The analysis that the total torque can be decomposed into a stationary component and a fast-varying component, 
with no significant coupling between both, assumes that differential Lindblad torque is not significantly altered by turbulence. 
If we apply the same tactics to the $\gamma=10^{-3}$ scenario, we would expect the timescale for the accumulative effect of the linear torque to be significant to be around $1200$ orbits. 
However, 
for such strong turbulence corresponding to $\langle\alpha_R\rangle \sim 0.04$, structure of density waves will be disrupted even after averaging the density distribution across orbits, 
similar to the $\langle\alpha_R\rangle \sim 0.1$ experiments in \citet{ChenLin2023}, 
which led us to believe that the running-average showing no sign of convergence towards the $\gamma=0$ value is genuine evidence that random walk dominated this chaotic migration process. 
In support of this claim, 
we highlight the torque density distribution of the $\gamma= 10^{-3}$ in Figure \ref{fig:2d}. The top panel represents the torque density of the $\gamma=0$ model at 2000 orbits in which the planet has 
migrated to $r \simeq 0.9 r_0$. In this classical flow pattern picture, 
the main contribution of Lindblad torque is associated with the competition between the leading waves extending towards the upper left (positive torque) and the trailing waves extending towards the lower right (negative torque) \citep{Tanaka2002}.
The middle panel is the snapshot of the $\gamma=10^{-3}$ model in which the planet's
orbit has not evolved significantly due to the turbulent interruption of the planet's tidal wake. 
The bottom
panel is a time average (over 200 orbits) torque density (for the $\gamma=10^{-3}$ model) which reduce the 
fluctuation but still show no sign of the density-wave structure emerging and is generally smaller than magnitude than the inviscid case.

% FFFFFFFFFFFFFFFFFFFFFF %
% FFFFFFFFFFFFFFFFFFFFFF %

% FFFFFFFFFFFFFFFFFFFFFF %
% FFFFFFFFFFFFFFFFFFFFFF %

% FFFFFFFFFFFFFFFFFFFFFF %
% FFFFFFFFFFFFFFFFFFFFFF %

% FFFFFFFFFFFFFFFFFFFFFF %
\begin{figure}
\centering
\includegraphics[width=0.99\hsize]{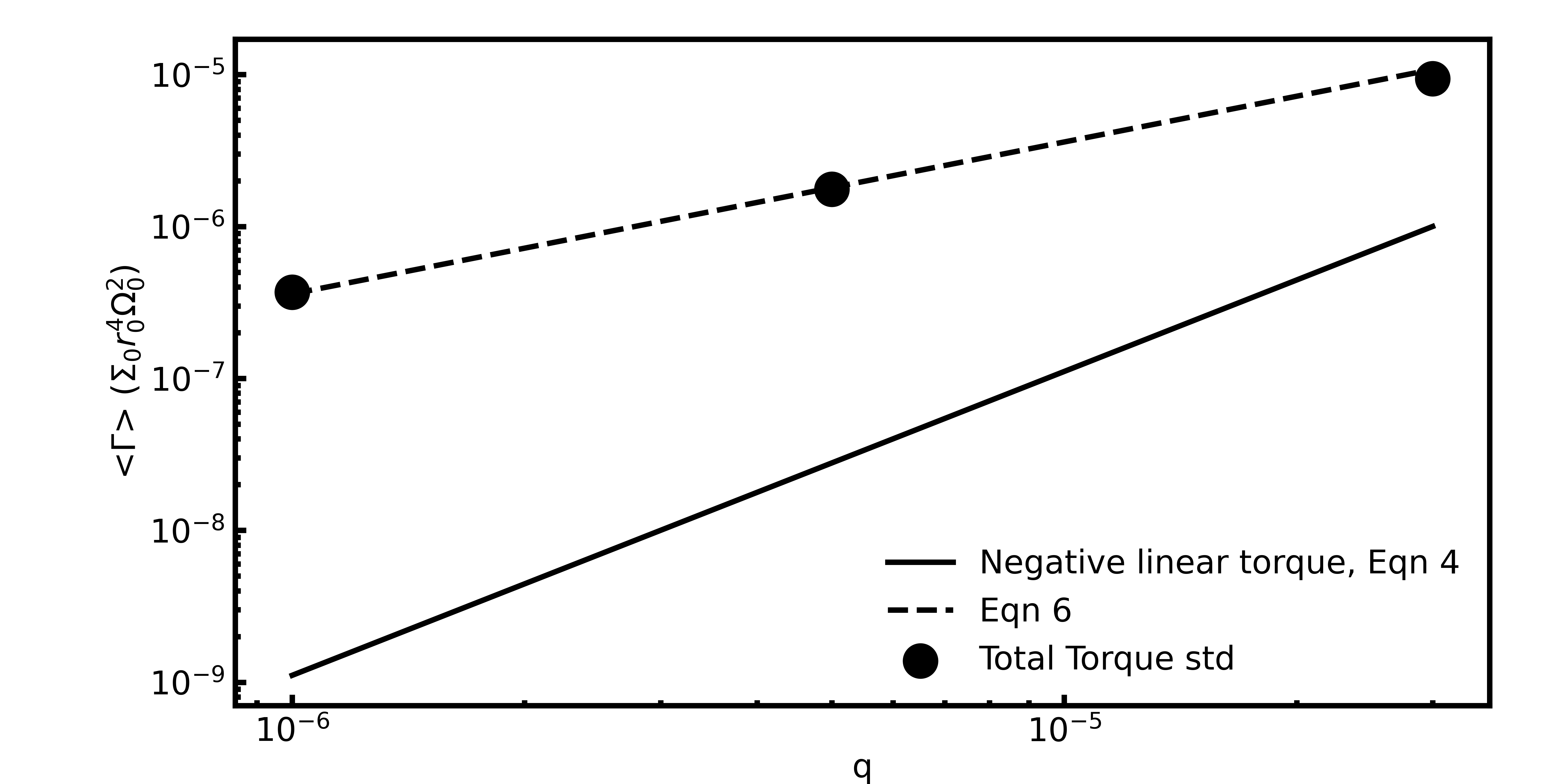}
\caption{Black symbols indicate the standard deviation of the migration torque $ \langle \Gamma \rangle $ measured over 500-2000 orbits for different values of mass ratio $q$. The values are closely fitted by Eqn \ref{eqn:dispersion} (dashed line). The magnitude of the negative linear torque (Eqn \ref{eqn:classic}) with $C_L = 1$ is plotted with a solid line.
}\label{fig:summary_q}
\end{figure}
% FFFFFFFFFFFFFFFFFFFFFF %

% FFFFFFFFFFFFFFFFFFFFFF %
\begin{figure}
\centering
\includegraphics[width=0.99\hsize]{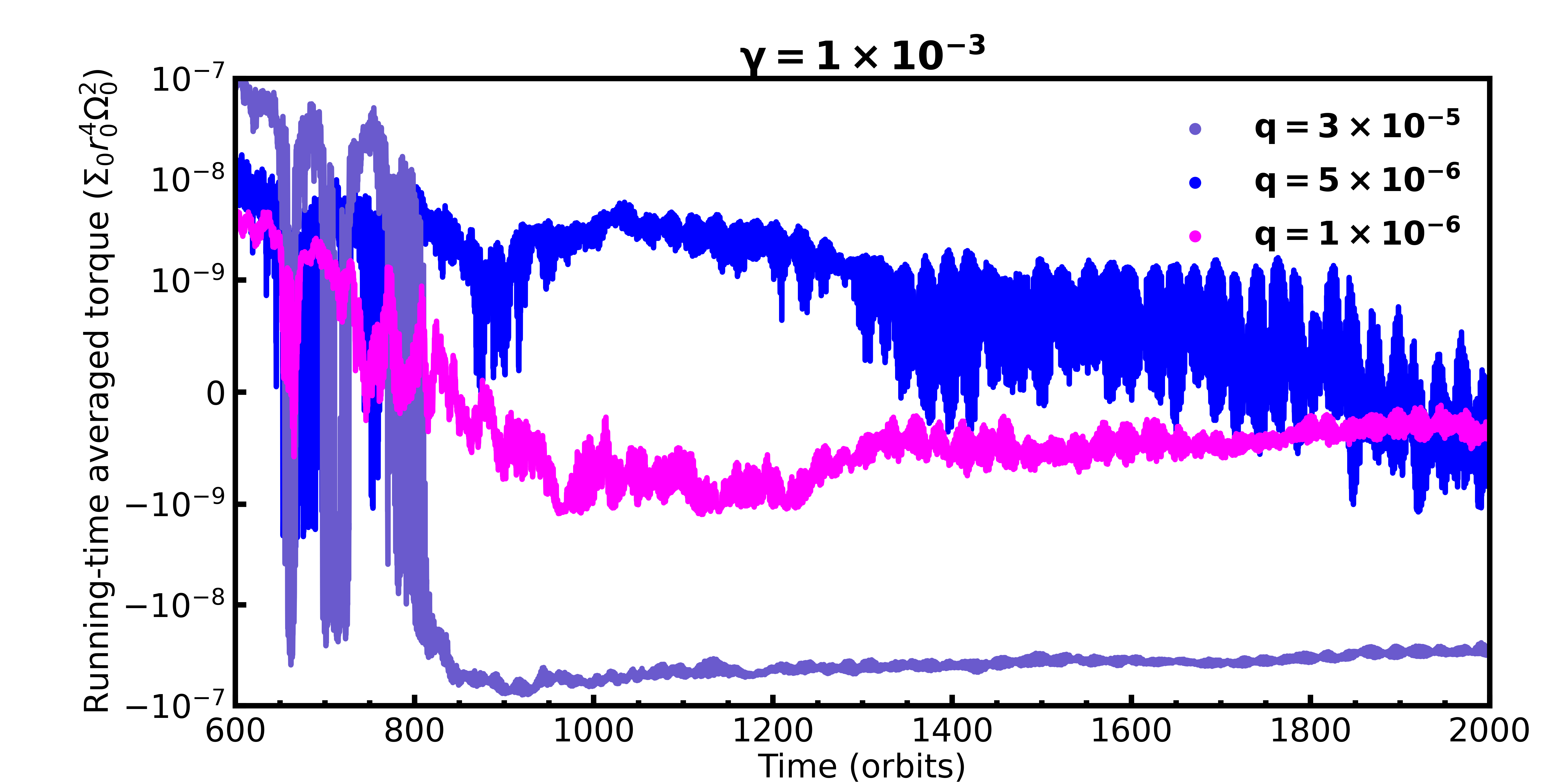}
\caption{Same as Figure \ref{fig:rtatorque} but used to compare different planet-to-star mass ratio $q$ with a fixed $\gamma=1\times10^{-3}$. No truncation of turbulent modes are performed. }\label{fig:q}
\end{figure}
% FFFFFFFFFFFFFFFFFFFFFF %

\subsection{Dependence on Planet Mass}

% FFFFFFFFFFFFFFFFFFFFFF %
\begin{figure}
\centering
\includegraphics[width=0.99\hsize]{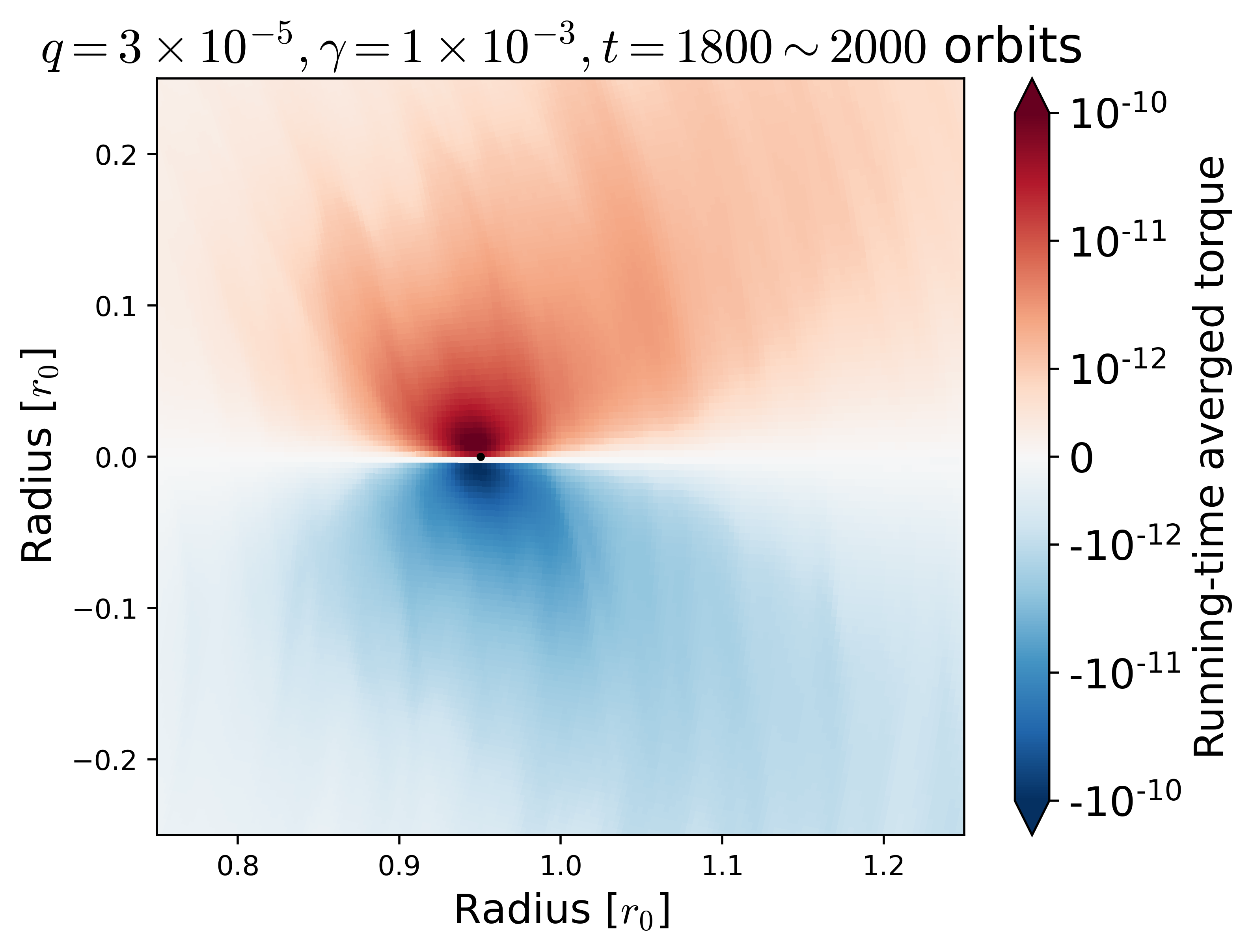}
\caption{Same as the bottom panel in Figure \ref{fig:2d} but for $q = 3\times10^{-5}$ with $\gamma=1\times10^{-3}$.}\label{fig:2d_q}
\end{figure}
% FFFFFFFFFFFFFFFFFFFFFF %
 
In this section, we further compare the torque evolution of the 
fiducial case (with $q = 5\times10^{-6}$) as well as companion mass ratios $q = 10^{-6}$ and 
$q = 3\times 10^{-5}$. 
The scale height $h_0=0.03$ is unchanged and all azimuthal wavenumbers are included. 
In Figure \ref{fig:summary_q} we plot the standard deviation of the torque (normalized by $\Sigma_0 r_0^4 \Omega_0^2$)in these 
three cases over 500-2000 orbits, 
fitted by 

\begin{equation}
 \langle \Gamma \rangle \approx 3.6\times 10^2 q \gamma \Sigma_0 r_0^4 \Omega_0^2, \ \ \ \ \text{or} \ \ \ \ \langle C_{\rm turb} \rangle =  3.6\times 10^2 q^{-1} h_0^2 \gamma 
 \label{eqn:dispersion}
\end{equation}
conforming with Eqn 10 of \citet{BaruteauLin2010} with a factor of \%50 discrepancy.

Figure \ref{fig:q} shows the evolution of the running-average torque for three different mass ratios under strong turbulence, also normalized by $\Sigma_0 r_0^4 \Omega_0^2$.
%the $\gamma=0$ inviscid result is also plotted in dotted lines correspondingly, and is $\sim q^{-2}$ smaller than the fiducial case at $\gamma=0$ in magnitude, conforming with \citet{Paardekooper2010} linear predictions. 
For the low mass ratio $q=10^{-6}$ case, the profile settles towards a fluctuating torque with an amplitude smaller than the linear torque, 
which is predicted to be $\sim 10^{-9}$ in corresponding units $(\Sigma_0 r_0^4 \Omega_0^2)$, 
suggestive for chaotic migration similar to the fiducial model. 
Nevertheless, since $|\langle C_{\rm turb} \rangle/C_{\rm L}|$ is larger than the fiducial case, 
we expect a robust convergence timescale of $\sim 2\times 10^4$ orbits,
only after which we can confidently tell if the linear background torque is completely suppressed. 
We did not run the simulation long enough due to computational constraints, 
but we comment that for very low masses, both the linear and fluctuating components are expected to be minimal, 
resulting in an overall small migration effect for all $\langle \alpha_{\rm R} \rangle$.

In the high mass case with marginal gap-opening $q\sim h_0^3 = 2.7 \times 10^{-5}$, 
the fluctuation $(\propto q)$ is larger than the linear torque $(\propto q^2)$ by only one order of magnitude now, 
suggesting it's much harder to drive chaotic migration of high-mass companions through 
turbulence \citep{BaruteauLin2010}. 
Naturally we do expect gap-opening planets' migration to be less perturbed by turbulent effects, as shown in MHD 
simulations \citep{Aoyama2023}.  
In Figure \ref{fig:q} we see the average of $q = 3\times 10^{-5}$ converging towards $\sim - 10^{-7}$, 
which suggest that long-term evolution of this gap opening companion will be linear torque dominated. From the specific torque distribution averaged over the last 200 orbits for $q = 3\times 10^{-5}$ (Figure \ref{fig:2d_q}), we also observe the emergence of density wave structure (trails extending towards the upper left and lower right) in contrast to the fiducial case (Figure \ref{fig:2d_q}).
Interestingly, the value of converged $\Gamma$ is still about one order-of-magnitude smaller than the expected linear torque from Eqn \ref{eqn:classic}.
This might be due to partial gap opening or non-linear enhancement of positive corotation torque, which is not the main focus of our study. 

%{\color{red} small inconsistency here is the convergence timescale for a linear torque of $10^{-7}$ and dispersion of $10^{-5}$ is 10000 orbits, so how did our running average converge after 100 orbits?}

Extrapolating to different values of $q$ and $\gamma$ or $\langle \alpha_R \rangle$, 
we generally expect a critical transition $q$ for a given level of turbulence $\alpha_R$ above which the planet shifts from chaotic migration towards linear migration. 
Meanwhile, 
the mean value of $\langle \Gamma \rangle$ begins to become significant. If the marginal gap opening case is close to linear migration. 
We can determine for $\langle \alpha_R \rangle = 0.04, h_0=0.03$ 
the transition $5\times 10^{-6} \lesssim q_{\rm crit} \lesssim 3\times 10^{-5}$. 
The same paradigm can be applied to a subsequent parameter survey to conclude a comprehensive scaling for $q_{\rm crit}$. 

{\citet{Chen2022} investigated the critical planet orbital eccentricity above which circum-companion flow patterns are disrupted. They concluded that when the characteristic epicyclic impact velocity $e \Omega_0 r_0$ is large, the background Keplerian shear would dominate over the classical flow pattern down to the modified Bondi radius $R_{\rm B, ecc} \sim GM_p/(e \Omega_0 r_0)^2$. 
Analogously, we can introduce a ``turbulent" Bondi radius of the turbulent velocity based on the turbulent velocity amplitude $\Delta v$, where by definition of Reynolds stress $\Delta v^2/c_s^2 \sim \langle \alpha_{\rm R} \rangle$, and assume on a time-average sense turbulent motions can  strongly dominate down to the scale of $R_{\rm B, turb} \sim  GM_p/\Delta v^2$ from the planet.}

%\begin{equation}
 %   R_{B, turb} = \dfrac{GM_p}{\langle \alpha_{\rm R} %\rangle c_s^2}
%\end{equation}
%which is a factor of $1/\langle \alpha_{\rm R} \rangle$ larger than the classic Bondi radius. 

{Since Lindblad resonances are effective at radial distances of a few pressure scale height $\sim H = h r$, we might expect strong chaotic fluctuation to significantly impact the type I torque when $R_{B, turb}$ is smaller than some factor $\lambda$ of $H$, 
{\it i.e.} when ${q} \lesssim  \lambda \langle \alpha_{\rm R} \rangle h^3$.}

%\begin{equation}
 %   \dfrac{q}{\langle \alpha_{\rm R} \rangle  h^2 } >   \lambda h, \ \ \ \ {\rm i.e.} \ \ \ \ q/h^3 > 
%    \lambda \langle \alpha_{\rm R} \rangle.
%\end{equation}

Here $4 < \lambda < 25$ is loosely constrained from our limiting test cases. 
More accurate values need to be obtained by extensive simulations. Such a determination is technically  
straightforward since the transition from classical to chaotic migration can be distinctively identified 
by plotting the running time average of torque over long enough timescales. 

% ======================================= %
\section{Conclusions}\label{sec: conclusion}
% ======================================= %

In this letter, we use hydrodynamical simulations with a driven turbulence prescription to show that low mass planets, or generally disc-embedded companions migrating in strongly turbulent discs are subject to highly stochastic driving forces, 
and the time-average residue migration torque becomes negligible after a few thousand orbits. 
For a $q = 5\times 10^{-6}$ companion, when the $\langle \alpha_R \rangle \sim 4\times 10^{-4}$, the average Lindblad torque and corotation torque will dominate after the decay of the running-average of turbulent torque component, 
as in \citet{BaruteauLin2010}, 
and become subject to linear expectations. 
However, 
for stronger turbulence corresponding to $\langle \alpha_{\rm R} \rangle \sim 4\times 10^{-2}$, 
the classical resonances are disrupted and the residue for running-time average does not tend to saturate to the classical predictions of Type I torque. 
This randomizing effect is stronger for lower companion mass, and becomes less significant for marginally gap-opening companion mass.

This stochastic migration paradigm applies to planet migrating in outer PPDs dominated by strong gravito-turbulence, 
as well as sBHs in AGN discs under influence of gravito-turbulence and the density waves generated by the swarm of embedded objects themselves \citep{Goodman2001}, 
including intermediate mass sBHs and massive stars \citep{Cantiello2021}. The $\langle \alpha_R \rangle \sim 4\times 10^{-2}$ requirement for our fiducial example appears to be reaching the upper limit of quasi-steady gravito-turbulence \citep{Gammie2001,Johnson2003,Deng2017}. 
However, even if intense cooling triggers disc fragmentation, 
the gas would not be completely deposited into fragments before the disc region in between the fragments re-stablish quasi-steady state at the maximum turbulence level, provided that the fragments formed do not play a major role in the energy and angular momentum budget of the disc. This self-regulation can either be due to a prolonged cooling timescale from lower surface density (more relevant to gas pressure dominated PPDs), 
or auxiliary heating mechanisms such as large scale spiral shocks or star formation heating (more relevant to AGN discs) \citep{Goodman2003,SirkoGoodman2003,Thompson2005}. 
Moreover, lower mass companions, whose flow patterns are more susceptible to disruption, would have lower transition turbulence parameters which widens the viable viscosity parameter space for chaotic migration. 
In light of these considerations, we propose that chaotic migration can indeed operate in a wide range of parameter space relevant to AGN discs, as well as a significant range of parameter space relevant to PPDs.
Regarding the dependencies of the magnitude of the random torque $\langle C_{\rm turb} \rangle$, 
our measurement is consistent with the scaling summarised by \citet{BaruteauLin2010}, 
which can be recast in terms of $\langle \alpha_R \rangle$ as:

\begin{equation}
    \langle C_{\rm turb} \rangle \approx 360 q^{-1} h_0^2 \gamma \approx 60 q^{-1} h_0^3 \langle \alpha_{\rm R} \rangle^{1/2} 
\end{equation}
which serves as a quantitative formula for future population studies that examine the effect of random driving torque on AGN channel GW statistics. 
In terms of a reference gravitational stability parameter $Q_0 \equiv h_0 \Omega_0 ^2 / \pi G \Sigma_0$, 
the net effect of turbulent migration will move the companion radially by a fraction of

\begin{equation}
    {\Delta r \over r_0} = \dfrac{\langle \Gamma \rangle}{M_p r_0^2 \Omega_0 }\dfrac{2\pi}{\Omega_0} 
    = 3.8\times 10^2 \dfrac{\Sigma_0 r_0^2 }{M_p} q h_0  \langle \alpha_{\rm R} \rangle^{1/2} 
     \simeq {10^2 \over Q_0} \langle \alpha_{\rm R} \rangle^{1/2}  h_0^2 ,
\label{eq:randomwalk}
\end{equation}
in either the inward or outward direction over each orbital timescale  and its magnitude would increase
with time by an amount $\sim (\Omega_0 t/2 \pi)^{1/2}$.  This estimate (Eq. \ref{eq:randomwalk}) is consistent with the numerical
simulations (Figures \ref{fig:rtatorque} \& \ref{fig:2d}) which show much slower random-walk diffusion (with $\gamma=10^{-3}$)
rather monotonic migration (with $\gamma=0$).

To first order, we expect chaotic migration to prevent sBHs from monotonically migrating towards a migration trap where they become dynamically crowded, 
with many migrating outwards as well as inwards. 
Consequently, three-body interactions and hierarchical mergers in migration traps will be less frequent as previously predicted \citep{Tagawa2020}. 
Although the self gravity of the disc's is neglected in these simulation,  the scaling relation in Eqn. \ref{eq:randomwalk} indicates
that the extend of $\Delta r \lesssim r_0$ during the lifespan of massive stars \citep{alidib2023} in typical AGN discs with 
marginal gravitational stability \citep{paczynski1978, Goodman2003}, modest gravito-turbulence \citep{deng2020},
and small scale height \citep{starkey2023}.  This slow diffusion
enables the heavy element pollution of AGN disc \citep{Huang2023} by massive stars before they migrate into and consumed by
the SMBH {without the need}
%in the absence 
of universal migration traps \citep{Bellovary2016}.
It is worth noting that strong three-body interaction 
in migration traps has been invoked to explain the low mean effective spin seen in LIGO-Virgo observations \citep{LIGO2021-third2}, assuming a significant portion of which is indeed from the AGN channel.  On the other hand, \citet{Li2022,Chen2022,ChenLin2023} has proposed other natural ways from the local gas accretion to reduce the magnitude as well as the mean of effective sBH spin from gas accretion, 
which does not rely on frequent interactions between sBH and sBH binaries.

{We also propose a scaling for the transition companion mass ratio from classical to chaotic migration $q_{\rm crit} \propto \langle \alpha_{\rm R} \rangle h_0^3$, 
which can be further tested and refined
by conducting extensive parameter survey using numerical simulations.   The distinct nature of the transition allows for straightforward testing. 
The parameter survey can be performed using either a comparable code setup or by incorporating more realistic turbulence treatments in subsequent works.}

%A final comment is that one can estimate the eddy-turnover factor \citep{ChenLin2023}, which is the number of eddy-turnover timescales/correlation timescales within a typical disc lifetime, to be 

%\begin{equation}
%    \mathcal{S} =     \left\{
%    \begin{aligned} &    3\times 10^4 \left(\frac{M_{*}}{ M_\odot} \right)^{3/2}  \left(\frac{\dot{M}}{ 10^{-6 }M_\odot/{\rm yr}} \right)^{-1} \left(\frac{r_0}{10 \rm AU} \right)^{-3/2} \text{ in PPDs}
 %   \\
%    &  3\times   10^4 \frac{\eta_\bullet}{0.1} \left(\frac{M_{\bullet}}{10^8 M_\odot} \right)^{1/2} \left(\frac{r_0}{\rm 0.3 pc} \right)^{-3/2}  \text{ in AGN discs}.
  %  \end{aligned}
%    \right.
%\end{equation}

%Here we assume the typical disc lifetime to be of order $M_*/\dot{M}$ or $M_\bullet/\dot{M}$ respectively, while $\dot{M}$ is the accretion rate onto the central object. The SMBH accretion rate is measured in terms of the Eddington efficiency factor $\eta_\bullet$. In either the gravito-turbulent outskirts of PPDs or AGN discs, this number will be quite large, and chaotic migration is expected to dominate initial conditions for the radial distribution of embedded low-mass companions over the disc evolution timescales.

% ======================================= %
\section*{Acknowledgements}
% ======================================= %
We thank Cl\'ement Baruteau for helpful discussions and sharing the turbulent module of FARGO and Ya-Ping Li for his guidance on data analysis. We thank Sergei Nayakshin, Richard Nelson, Kevin Schlaufman, Xing Wei and the anonymous referee for inspiring discussions or comments. This research used DiRAC Data Intensive service at Leicester, operated by the University of Leicester IT Services, which forms part of the STFC DiRAC HPC Facility (\href{www.dirac.ac.uk}{www.dirac.ac.uk}).

% ======================================= %
\section*{Data availability}
% ======================================= %

The data obtained in our simulations can be made available on reasonable request to the corresponding author. 



\bibliographystyle{mnras}
\bibliography{turb-disk} % if your bibtex file is called example.bib

%%%%%%%%%%%%%%%%%%%%%%%%%%%%%%%%%%%%%%%%%%%%%%%%%%

% Don't change these lines
\end{CJK*}
\bsp	% typesetting comment
\label{lastpage}
\end{document}